\documentclass{PoS}

\usepackage{todonotes}
\usepackage{amsmath}
\usepackage{cite}

\newcommand{\Psibar}{\overline{\Psi}}
\newcommand{\Oalpha}{O(\alpha)}
\newcommand{\Oalphasquare}{O(\alpha^2)}

\title{Electromagnetic Corrections to Meson Masses and the HVP}

\ShortTitle{Electromagnetic Corrections to Meson Masses and the HVP}

\author{\speaker{Vera G\"ulpers\footnote{Speaker}, James Harrison}, Andreas 
J\"uttner, Christopher Sachrajda\\
        School of Physics and Astronomy, University of Southampton\\
        Southampton SO17 1BJ, United Kingdom\\
        E-mail: \email{\{V.M.Guelpers, J.Harrison\}@soton.ac.uk}}
\author{Peter Boyle, Antonin Portelli\\
School of Physics and Astronomy, University of Edinburgh\\
Edinburgh EH9 3JZ, United Kingdom
}        
\author{RBC and UKQCD Collaborations}

\abstract{We present an exploratory study of the electromagnetic 
corrections to meson masses and the hadronic vacuum polarization using  
$N_f=2+1$ Domain Wall fermions. These corrections are estimated with two 
different approaches, a stochastic approach using $U(1)$ gauge configurations 
for the photon fields, and a perturbative approach through a QED perturbative 
expansion of the QCD+QED path integral. We compare results and statistical 
errors from both methods.}

\FullConference{34th annual International Symposium on Lattice Field Theory\\
		24-30 July 2016\\
		University of Southampton, UK}

\begin{document}

\section{Introduction}
In recent years the computation of many quantities relevant to Standard Model 
phenomenology in Lattice QCD has reached 
a precision of $\lesssim1\%$ \cite{FLAG}. However, such calculations are 
usually done in the 
isospin symmetric limit $m_u=m_d$ and without including QED effects, i.e.\ 
considering quarks as being electrically neutral. Both effects can be of the 
order of 
$1\%$ \cite{antonin_lat14} and need to be included when aiming for 
calculations at this level of precision. 
\par
The treatment of electromagnetic effects in lattice calculations requires the 
inclusion of QED in the path integral. The expectation value of a quantity $O$
is then given in the Euclidean path integral formalism by
\begin{equation}
  \left<O\right> = \frac{1}{Z}\! \int 
\!\!\mathcal{D}[U]\,\mathcal{D}[A]\,\mathcal{D}[\Psi,\Psibar]\,\,O\,\,
e^ { -S_F[\Psi,\Psibar ,A, U]}\,\,e^{ - S_\gamma[A]}\,e^{ - S_G[U]}\,,
\label{eq:pathintegral}
\end{equation}
with photon fields $A$, the $SU(3)$ gauge fields $U$ and the quark fields 
$\Psi$ and $\Psibar$. The fermionic action 
$S_F[\Psi,\Psibar ,A, U]$ now also includes couplings of quarks to photons.
We follow two approaches to include QED in our lattice calculations. One is a 
non-perturbative treatment of QED, where $U(1)$ photon fields are generated 
stochastically. We generate the $U(1)$ photon fields independently 
of $SU(3)$ gauge fields. In 
this electro-quenched approximation the sea quarks are electrically neutral.
The $U(1)$ photon fields are then multiplied with the $SU(3)$ gauge 
links 
\cite{firststoch,Blum2007,RBCUKQCD_qed,BMW_PRL,BMW_science,BMW_quarkmass,
QCDSF_JHEP,QCDSF_JPhysG,QED_Milc}
\begin{equation}
U_\mu(x)\rightarrow e^{ieA_\mu(x)} U_\mu(x)\,.
\label{eq:u3links}
\end{equation}
In the following we will refer to this method as the 
\textit{stochastic method}. 
\par
On the other hand QED can also be treated perturbatively by expanding the 
path integral (\ref{eq:pathintegral}) as a series in the fine structure 
constant 
$\alpha=e^2/4\pi$ \cite{pert_rome}
\begin{equation}
 \left<O\right> = \left<O\right>_{e=0} + 
\frac{1}{2}\,e^2\frac{\partial^2}{\partial 
e^2}\left<O\right>_{e=0} + \Oalphasquare\,.
\label{eq:eexpansion}
\end{equation}
The expectation values $\left<\cdot\right>_{e=0}$ on the right hand side of 
equation (\ref{eq:eexpansion}) are then calculated in pure QCD. We will refer 
to this method as the \textit{perturbative method} in the following.
\par%
Strong isospin breaking effects are taken into account by putting different 
values for up- and down-quark masses in the valence sector. We note that a more 
comprehensive method for the treatment of strong isospin breaking is to expand 
the path integral in the isospin symmetric quark mass as proposed in
\cite{strongIB_expansion}.
\par
A more detailed description of the stochastic and the perturbative method 
is given in sections \ref{sec:stochastic} and \ref{sec:perturbative}. In 
sections \ref{sec:mesonmasses} and \ref{sec:hvp} we will show results for the 
QED correction to meson masses and the hadronic vacuum polarization, 
respectively and compare results and statistical errors from both approaches.

For this exploratory study we use a $64\times24^3$ lattice with an 
inverse lattice spacing of $a^{-1}=1.78$~GeV and $N_f=2+1$ dynamical flavors of 
Domain Wall Fermions \cite{RBCUKCD_DWF}. The pion mass (without QED) is 
$\approx340$~MeV. We use 
different masses for the valence up and down quarks, such that their difference 
approximately corresponds to the physical mass difference 
\cite{BMW_quarkmass} and we use the physical value for the valence strange 
quark mass \cite{RBCUKQCD_ensemble}. We use the same bare-quark masses when 
caluclating with or without QED.
\section{Stochastic Method}
\label{sec:stochastic}

In the electro-quenched approximation, $U(1)$ photon fields are generated
independently of $SU(3)$ gauge fields. We define the lattice $U\left(1\right)$
vector potential at mid-links, and denote this as
$A_{\mu}\left(x\right)\equiv A\left(x+a\hat{\mu}/2\right)$. We define the
non-compact photon action as:
\begin{equation}
	S_{\gamma}\left[A_{\mu}\right]=\frac{a^{4}}{4}\sum_{x}\sum\limits_{\mu,\nu}\left(\partial_{\mu}A_{\nu}\left(x\right)-\partial_{\nu}A_{\mu}\left(x\right)\right)^{2},\label{eq:lattice_EM_action}
\end{equation}
where $\partial_{\mu}f\left(x\right)=a^{-1}\left[f\left(x+a\hat{\mu}\right)-f\left(x\right)\right]$
is the forward derivative. Using integration by parts, this can be
written as
\begin{equation}
	S_{\gamma}\left[A_{\mu}\right]=\frac{a^{4}}{2}\sum_{x}\sum\limits_{\mu,\nu}A_{\mu}\left(x\right)\left(\partial_{\mu}^{*}\partial_{\nu}-\delta_{\mu\nu}\partial^{2}\right)A_{\nu}\left(x\right).
\end{equation}
where $\partial_{\mu}^{*}f\left(x\right)=a^{-1}\left[f\left(x\right)-f\left(x-a\hat{\mu}\right)\right]$
is the backward derivative and
$\partial^{2}\equiv\sum_{\mu}\partial_{\mu}^{*}\partial_{\mu}$. This action is
invariant under constant shift transformations
$A_{\mu}\left(x\right)\rightarrow A_{\mu}\left(x\right)+c_{\mu}$ and gauge
transformations $A_{\mu}\left(x\right)\rightarrow A_{\mu}\left(x\right)-\partial_{\mu}f\left(x\right)$.

The shift symmetry can be eliminated by removing the zero-mode of
the photon field. We choose to remove the zero-mode using the
$\mathrm{QED}_{L}$ formulation \cite{QED_L}, in which the spatial zero-mode is
removed on each time slice:
\begin{equation}
	a^{3}\sum_{\vec{x}}A_{\mu}\left(x_0,\vec{x}\right)=0\text{ for all }\mu,x_{0}.
\end{equation}
This condition is the most minimal subtraction that conserves the reflection
positivity of the theory with periodic boundary conditions \cite{BMW_science}.
The Feynman gauge can be imposed by adding a term to the action:
\begin{eqnarray}
	S_{\gamma,\mathrm{Feyn.}}\left[A_{\mu}\right] & = & S_{\gamma}\left[A_{\mu}\right]+\frac{1}{2}\sum_{x}\left(\sum_{\mu}\partial_{\mu}A_{\mu}\left(x\right)\right)^{2}\nonumber \\
	 & = & -\frac{a^{4}}{2}\sum_{x}\sum_{\mu}A_{\mu}\left(x\right)\partial^{2}A_{\mu}\left(x\right).
\end{eqnarray}
In momentum space, this action becomes
\begin{equation}
	S_{\gamma,\mathrm{Feyn.}}\left[A_{\mu}\right]=\frac{1}{2N}\sum\limits_{k,\vec{k}\neq0}\hat{k}^{2}\sum_{\mu}\left|\tilde{A}_{\mu}\left(k\right)\right|^{2},
\end{equation}
where $\hat{k}_{\mu}=\frac{2}{a}\sin\left(ak_{\mu}/2\right)$ and $N$ is the
number of lattice sites. All elements of the photon field
$\tilde{A}_{\mu}\left(k\right)$ are therefore independent, and a momentum-space
QED gauge configuration can be generated by drawing each value
$\tilde{A}_{\mu}\left(k\right)$ from a Gaussian distribution with variance
$2N/\hat{k}^{2}$.

In this work, we use QED gauge configurations in both Feynman and
Coulomb gauges. The Coulomb gauge fixing condition
$\sum_{j}\partial_{j}^{*}A_{j}\left(x\right)=0$, where the index $j$ runs over
the spatial directions, defines an operator which projects a field
configuration into Coulomb gauge \cite{BMW_science}:
\begin{eqnarray}
	A_{\mu} & \rightarrow & A_{\mu}'=\left(P_{C}\right)_{\mu\nu}A_{\nu}\\
	\left(P_{C}\right)_{\mu\nu} & = & \delta_{\mu\nu}-\left|\vec{\hat{k}}\right|^{-2}\hat{k}_{\mu}\left(0,\vec{\hat{k}}\right)_{\nu}.
\end{eqnarray}

After generating QED gauge configurations in momentum space, these
are transformed into position space using an inverse Fast Fourier Transform.
Since the photon field is a real vector field, the imaginary part of
$A_{\mu}\left(x\right)$ is set to zero after the Fourier transform.

Once position-space photon field configurations are generated, these
are multiplied with the $SU(3)$ gauge links as in (\ref{eq:u3links})
and the calculation of hadronic observables proceeds as in the case
without QED. This method allows the calculations of QED corrections up
to all orders within the electro-quenched approximation.

We remove $\mathcal{O}(e)$ noise by averaging
expectation values evaluated with electromagnetic couplings $+e$
and $-e$, thereby restoring exact charge-conjugation symmetry on
each configuration \cite{Blum2007}. We find that this process greatly
improves the signal-to-noise ratio, reducing statistical error by
an order of magnitude in some cases.

To calculate electromagnetic corrections, we compute correlators both
with and without QED. The stochastic method therefore requires three
inversions per quark flavour per source position; for electromagnetic
couplings $0,+e,-e$.

\section{Perturbative Method}
\label{sec:perturbative}
\begin{figure}[h]
\centering
\includegraphics[width=0.95\textwidth]{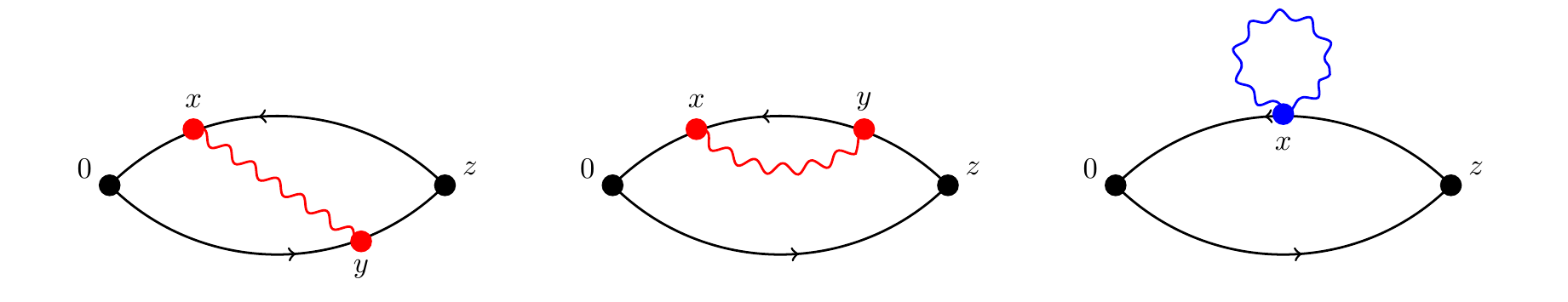}
 \caption{The three diagrams for the leading order QED corrections to mesonic 
two-point functions. Red and blue vertices represent insertions of the 
conserved vector current and the tadpole operator, respectively. From left to 
right the diagrams are photon exchange, self energy and tadpole.}
 \label{fig:qeddiagrams} 
\end{figure}

In a perturbative expansion of the path integral (\ref{eq:pathintegral}) in the 
fine structure constant $\alpha$ one obtains at lowest order diagrams with 
either two insertions of the conserved vector current or one insertion of the 
tadpole operator \cite{pert_rome}. This results in three different 
types of connected diagrams (photon exchange, self energy and tadpole) which 
are 
shown in figure \ref{fig:qeddiagrams}. Currently, we neglect quark-disconnected 
diagrams, which corresponds to quenched QED, i.e.\ neglecting electromagnetic 
effects for the sea quarks.   

To give an example, the photon exchange diagram for a charged kaon is given by 
a correlation function of the form
\begin{equation}
C(z_0) =\!\! 
\sum\limits_{\vec{z}}\!\sum\limits_{x,y}\!\textrm{Tr}\!\left[S^s(z,
x)\,\Gamma^c_\nu\,S^s(x,0)\,\gamma_5\,S^{u}(0,y)\,\Gamma^c_\mu\, 
S^{u}(y,z)\,\gamma_5\right]\Delta_{\mu\nu}(x-y)\,,
\label{eq:exch_kaon}
\end{equation}
where $S^f(x,y)$ denotes a quark propagator from $x$ to $y$ for a quark of 
flavor $f$, $\Delta_{\mu\nu}(x-y)$ is the photon propagator from $x$ to $y$ and 
$\Gamma^c_\mu$ stands for an insertion of the conserved vector current. In 
Feynman gauge, the photon propagator is given by
\begin{equation}
  \Delta_{\mu\nu}(x-y) = 
\delta_{\mu\nu}\,\frac{1}{V}\sum\limits_{k,\vec{k}\neq0}\,\,
\frac{e^{ik\cdot(x-y)}}{\hat{k}^2}\,,
\label{eq:photonprop}
\end{equation}
with the volume of the lattice $V=L^3\cdot T$.
In equation (\ref{eq:photonprop}) we have subtracted all spatial zero modes 
$\vec{k}=0$ of the photon propagator, which corresponds to the formulation of 
QED$_L$ \cite{BMW_science}.
\par
To compute correlation functions such as (\ref{eq:exch_kaon}), we rewrite the 
photon propagator as
\begin{equation}
\Delta_{\mu\nu}(x-y) =
\sum_u\Delta_{\mu\nu}(x-u)\left<\eta_i(u)\eta_i^\dagger(y)\right>_\eta = 
 \left<\tilde\Delta_{\mu\nu}(x)\eta^\dagger(y)\right>_\eta
\end{equation}
with stochastic sources $\eta$ that have the property
\begin{equation}
 \left< \eta_i(u)\eta_i^\dagger(y)\right>_\eta =
\delta_{u,y}\,.
\end{equation}
In this work, we use $\mathbb{Z}_2$ sources, where we randomly pick the entries 
from $\left\{\frac{1}{\sqrt{2}}(\pm1\pm i)\right\}$.
The photon propagator $\tilde\Delta_{\mu\nu}(x)$ can be calculated using a Fast 
Fourier Transform. 
We then calculate sequential quark propagators with insertions of the 
conserved 
vector current and either $\tilde\Delta_{\mu\nu}(x)$ or $\eta^\dagger(y)$, 
such 
that we can construct the diagrams shown in figure \ref{fig:qeddiagrams}. In 
total this requires $17$ inversions per valence quark and source position for 
Feynman gauge photon propagators, where only diagonal terms $\mu=\nu$ 
contribute (cf. equation (\ref{eq:photonprop})). 
For other gauges, e.g.\ Coulomb 
gauge, where also off-diagonal terms $\mu\neq\nu$ are non-zero, more inversions 
are required.
\par
We use one stochastic $\mathbb{Z}_2$ source $\eta$ for the photon insertion per 
QCD 
gauge configuration and source position.

\section{QED correction to meson masses}
\label{sec:mesonmasses}
In the following we will determine the QED correction to the pion and kaon 
masses. 
Although our main interest is in calculating the QED correction to the 
hadronic vacuum polarization contribution to the anomalous magnetic moment of 
the muon, the QED correction to the meson masses can serve as a valuable cross 
check of our calculation and as a starting point in comparing the stochastic 
and the perturbative approaches. 
\par
An independent calculation of the QED correction to meson masses on the same 
gauge ensemble has been previously done in \cite{RBCUKQCD_qed} using a 
stochastic 
approach. Those results can be used as a cross check for our implementation. 

\subsection{Results obtained using the Stochastic Method}

The mass of a meson determines the leading exponential decay of the
two-point function:
\begin{eqnarray}
C\left(t\right) & = & Ae^{-tm}+Ae^{-\left(T-t\right)m}+\dots\\
 & = & 2Ae^{-\left(T/2\right)m}\cosh\left(\left(t-T/2\right)m\right)+\dots 
\label{eq:twoptcosh}
\end{eqnarray}
where $T$ is the time extent of the lattice with periodic boundary conditions. 
The dots represent the contributions of excited states.
The mass can therefore be extracted by fitting a function 
of the form \eqref{eq:twoptcosh}
to the two-point function. The QED mass correction is then defined
as the correlated difference of the masses with and without QED:
\begin{equation}
\delta m=m_{\mathrm{QED}}-m_{0}\,.
\end{equation}
A suitable fit range can be determined from the plateau region of
the effective mass, defined by
\begin{equation}
\frac{C\left(t\right)}{C\left(t+1\right)}=\frac{\cosh\left(\left(t-T/2\right)m_{
\mathrm{eff}}\right)}{\cosh\left(\left(t+1-T/2\right)m_{\mathrm{eff}}\right)}\,.
\end{equation}

\begin{figure}[h]
 	\centering
	\includegraphics[width=0.48\textwidth]{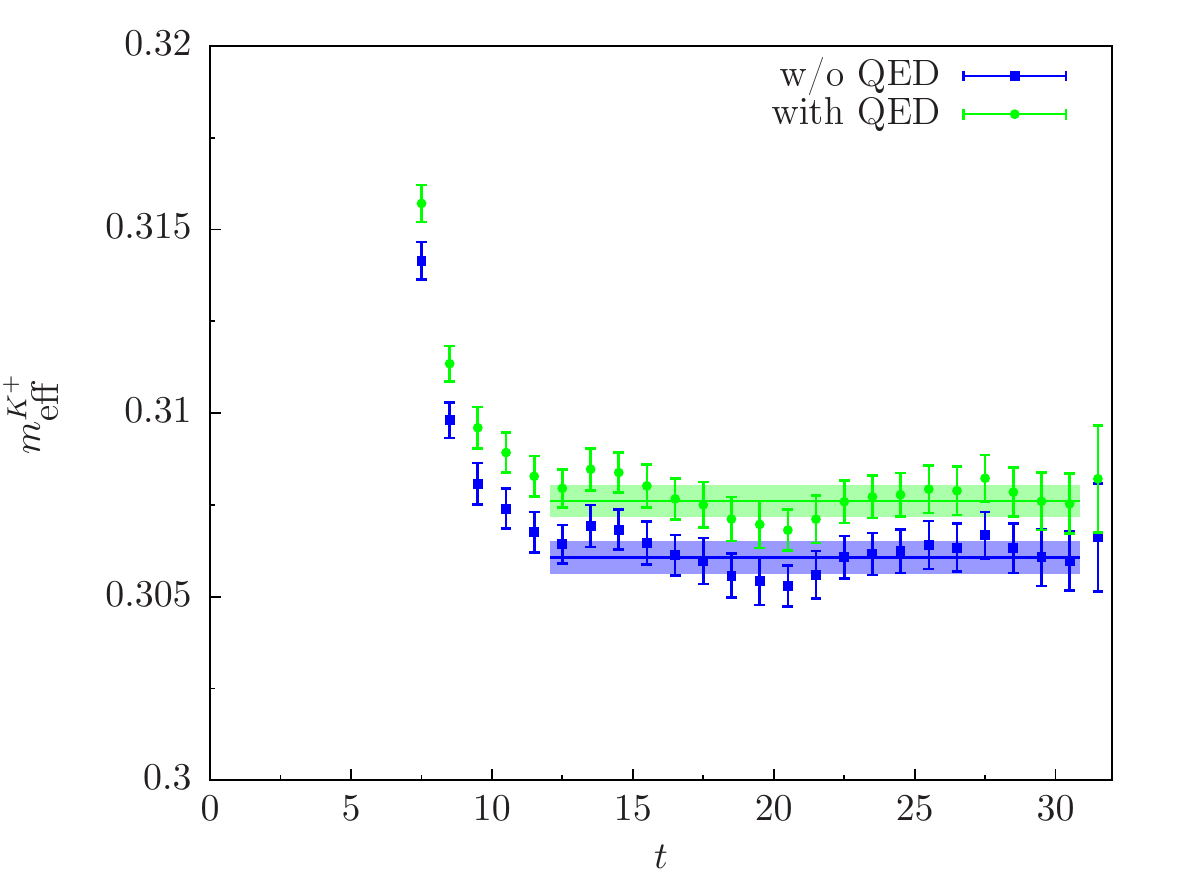}
	\includegraphics[width=0.48\textwidth]{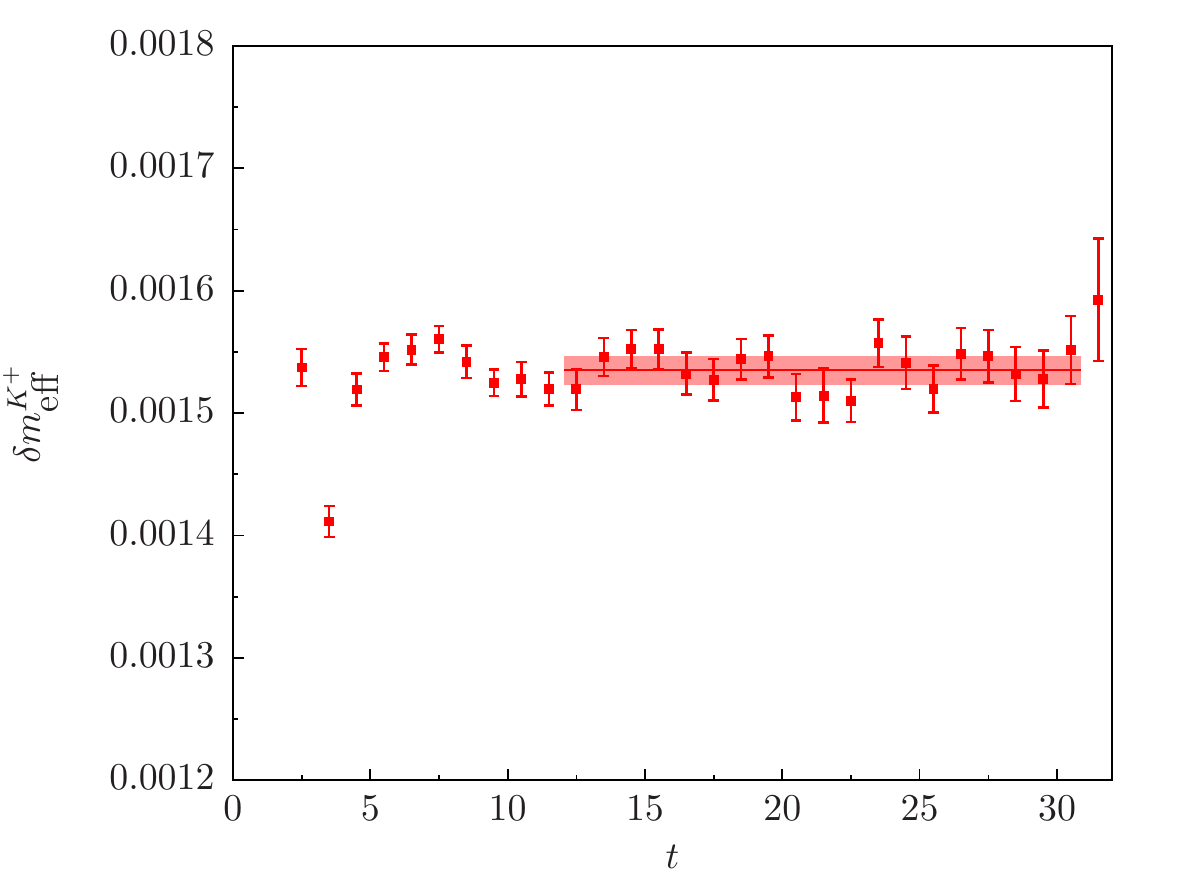}
	\caption{Effective mass of the charged kaon, along
	with the mass from a fit to the correlator, from the stochastic method
	in Coulomb gauge. The left plot shows the charged kaon mass with and
	without QED, and the right plot shows the QED mass correction $\delta 
m$.
	}
	\label{fig:stoch_kaon_mass}
\end{figure}

Figure \ref{fig:stoch_kaon_mass} shows the mass and QED mass correction
determined from a fit to the charged kaon correlator using the stochastic
method, along with the corresponding effective mass. These plots use
data from 87 configurations, with 16 $\mathbb{Z}_{2}$ wall sources per 
configuration,
and Coulomb-gauge $U\left(1\right)$ configurations.

\begin{figure}[h]
	\centering
	\includegraphics[width=0.48\textwidth]{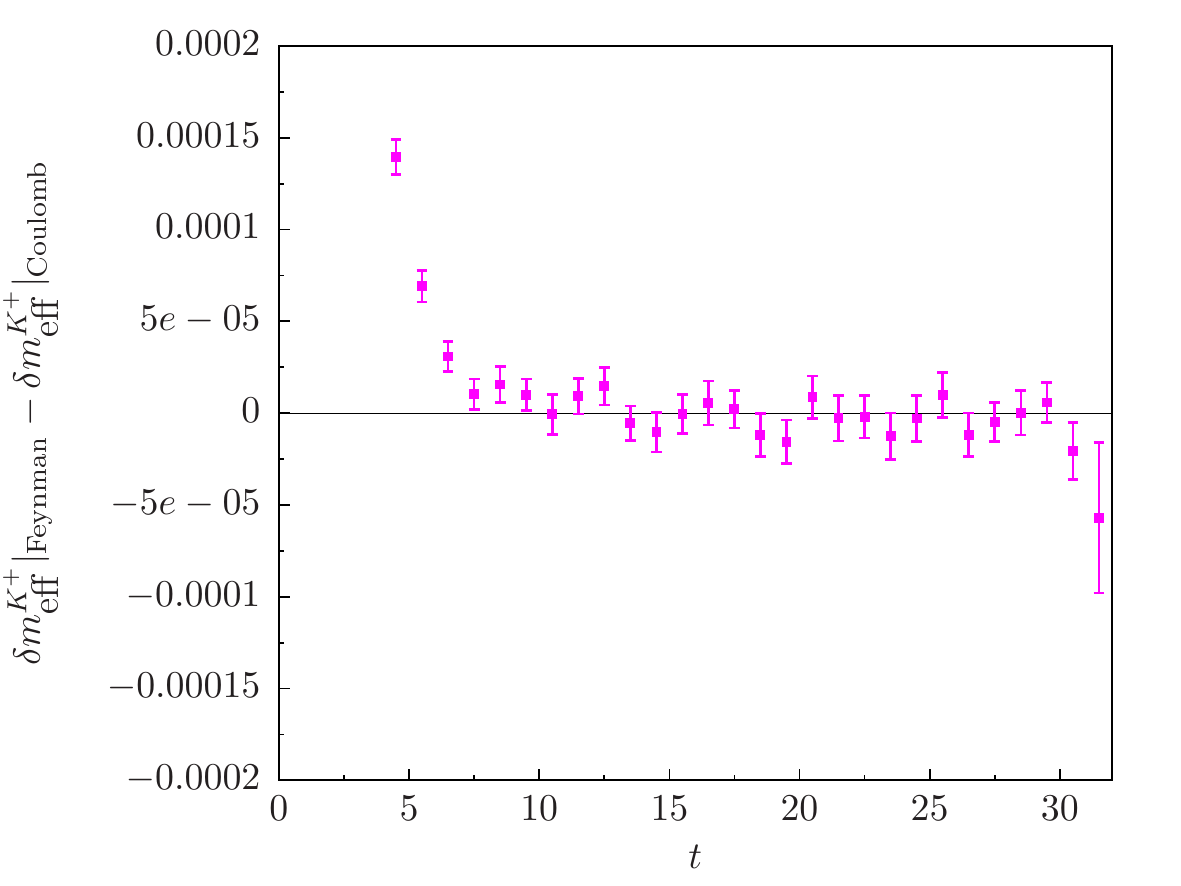}
	\caption{Correlated difference of the
	charged kaon effective mass QED correction evaluated in the Feynman
	and Coulomb gauges, using the stochastic method.}
	\label{fig:stoch_kaon_gauge_difference}
\end{figure}

We have calculated the QED mass correction using the stochastic method
on the same statistics with Feynman-gauge $U\left(1\right)$ configurations.
Figure \ref{fig:stoch_kaon_gauge_difference} shows the correlated
difference of the charged kaon effective mass QED correction evaluated
in the Feynman and Coulomb gauges. 
At early times, one can observe a
disagreement between the two gauges that
can be explained in the following way. In this
region the effective mass is contaminated by
excited states and in particular depends on
their creation amplitude, which is a gauge-dependent number. At later times, we 
see an expected agreement between the different gauges.

In table \ref{tab:cross-check} we compare our values for the pion
squared mass splitting $\Delta m^{2}=m_{\mathrm{QED}}^{2}-m_{0}^{2}$
with those from \cite{RBCUKQCD_qed}. Both analyses use a stochastic
method with Feynman-gauge $U\left(1\right)$ configurations, the same
$SU\left(3\right)$ gauge ensemble and the same fit range. We find
agreement between our work and the independent study at $1\sigma$.

\begin{table}[h]
	\centering
	\begin{tabular}{|c|c|c|c|}
		\hline 
		$q_{1}$ & $q_{2}$ & $\Delta m^{2}$ from this work & $\Delta 
m^{2}$ from \cite{RBCUKQCD_qed}\\
		\hline\hline
		2/3 & 2/3 & $\left(5.465\pm0.035\right)\times10^{-4}$ & 
$\left(5.406\pm0.064\right)\times10^{-4}$\\
		2/3 & -1/3 & $\left(7.677\pm0.052\right)\times10^{-4}$ & 
$\left(7.654\pm0.056\right)\times10^{-4}$\\
		-1/3 & -1/3 & $\left(1.341\pm0.009\right)\times10^{-4}$ & 
$\left(1.326\pm0.016\right)\times10^{-4}$\\
		\hline 
	\end{tabular}
	\caption{Comparison of pion squared mass splittings
	from this study, using the stochastic method in Feynman gauge, with
	those published in \cite{RBCUKQCD_qed}. Both quark masses are equal
	to the unitary light mass. Quark charges are in units of the electron
	charge $e$, and mass splittings are in lattice units.}
	\label{tab:cross-check}
\end{table}

Table \ref{tab:mesonmasses} contains the QED corrections to pion
and kaon masses from the stochastic method in Feynman gauge alongside results 
from the perturbative method. These
results have not been corrected for finite volume effects. An analytic formula 
to correct for finite volume effects in the meson masses is given in 
\cite{BMW_science}.

\subsection{Results obtained using the Perturbative Method}
\begin{figure}[h]
 \centering
 \includegraphics[width=0.55\textwidth]{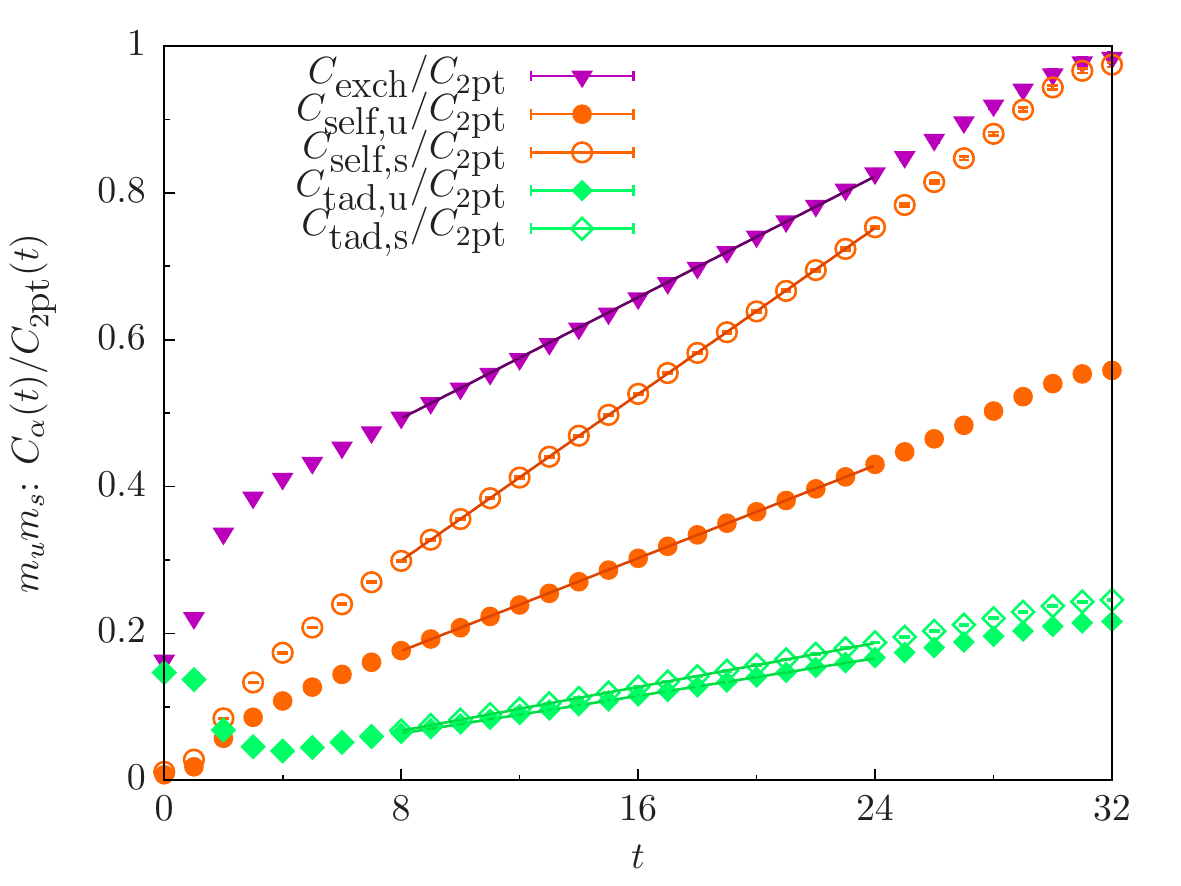}
 \caption{The QED correction to the mass of a charged kaon from the 
perturbative method. Orange circles and violet triangles are the contributions 
from the self energy 
and photon exchange diagram, respectively. The green diamonds are the 
contribution from the tadpole diagram. Where 
open and closed symbols are used, the open symbols refer to the photon being 
attached to the s quark, the closed symbols to the photon being attached to the 
u quark. Charge factors are not included in the data shown here.}
 \label{fig:qedmasscorr}
\end{figure}

The exponential decay of the two-point function including QED is determined 
by the mass $m_0$ without QED plus the leading order QED correction $\delta m$
\begin{equation}
 C(t) =  C_{0}(t) + C_{\Oalpha}(t) + \Oalphasquare = A\,e^{-(m_0+\delta m)\cdot 
t}\,
\label{eq:totalcorr}
\end{equation}
Expanding the exponent in (\ref{eq:totalcorr}) one finds that the QED 
correction to the meson mass can be estimated from the slope of the ratio of 
the QED correction $C_{\Oalpha}(t)$ to the two-point function and the two-point 
function $C_{0}(t)$ without QED \cite{pert_rome}
\begin{equation}
 \delta m = - \frac{\partial}{\partial t} \frac{C_{\Oalpha}(t)}{ C_{0}(t)}\,.
 \label{eq:deltam}
\end{equation}
Indeed we find a linear behavior in our data for $C_{\Oalpha}(t)/C_{0}(t)$ as 
can be seen in figure \ref{fig:qedmasscorr}. This plot shows the correlation 
functions of the three different types of QED correction divided by the 
two-point function without QED for a charged kaon. We have fitted a linear 
function to the data to obtain the slope, which can be related to the QED 
correction by (\ref{eq:deltam}).
\par
In table \ref{tab:mesonmasses} results for the QED correction 
to meson masses from the stochastic and the perturbative method are listed. The 
values are given 
in lattice units and have not been corrected for finite volume effects. 
Besides the QED corrections to the masses of charged and neutral pion as well 
as charged and neutral kaon, we quote the pion mass splitting 
$M_{\pi^+}-M_{\pi^0}$, which is given only by the photon exchange diagram, 
the contributions from the other diagrams cancel \cite{pert_rome}. Note, that 
we do not include the disconnected diagram for the neutral pion.
\par
\begin{table}[h]
 \centering
 \begin{tabular}{|c|c|c|}
   \hline
   quantity & stochastic & perturbative\\
   \hline\hline
   $aM^\gamma_{\pi^+}$ &$\left(1.968\pm0.013\right)\times10^{-3}$& 
$\left(1.963\pm0.016\right)\times10^{-3}$\\
   $aM^\gamma_{\pi^0}$ &$\left(0.885\pm0.007\right)\times10^{-3}$& 
$\left(0.877\pm0.016\right)\times10^{-3}$\\
   $a\left(M_{\pi^+}-M_{\pi^0}\right)$ 
&$\left(1.148\pm0.013\right)\times10^{-3}$& 
$\left(1.115\pm0.019\right)\times10^{-3}$\\
\hline
   $aM^\gamma_{K^+}$  & $\left(1.532\pm0.011\right)\times10^{-3}$&  
$\left(1.513\pm0.013\right)\times10^{-3}$\\
   $aM^\gamma_{K^0}$  & $\left(0.310\pm0.002\right)\times10^{-3}$& 
$\left(0.309\pm0.004\right)\times10^{-3}$\\
\hline
 \end{tabular}
 \caption{Results for the QED corrections to the meson masses in lattice units 
from the stochastic and perturbative method without finite volume correction.}
\label{tab:mesonmasses}
\end{table}

In addition, we have calculated the QED corrections to the meson masses 
with the perturbative method using the Coulomb gauge on a subset of the 
statistics 
and find agreement between the results in both gauges.

\subsection{Comparison of both methods}
\begin{figure}[h]
 \centering
\includegraphics[width=0.48\textwidth]{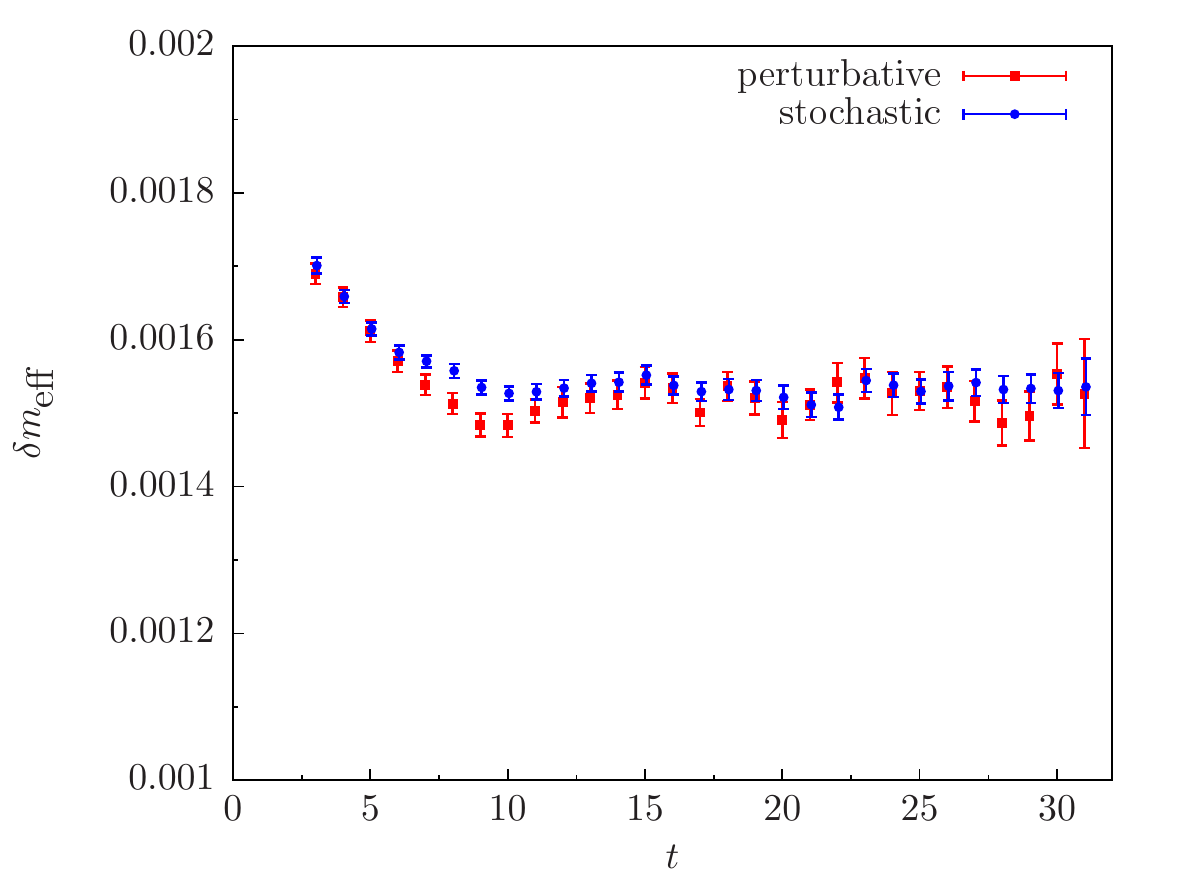}
\includegraphics[width=0.48\textwidth]{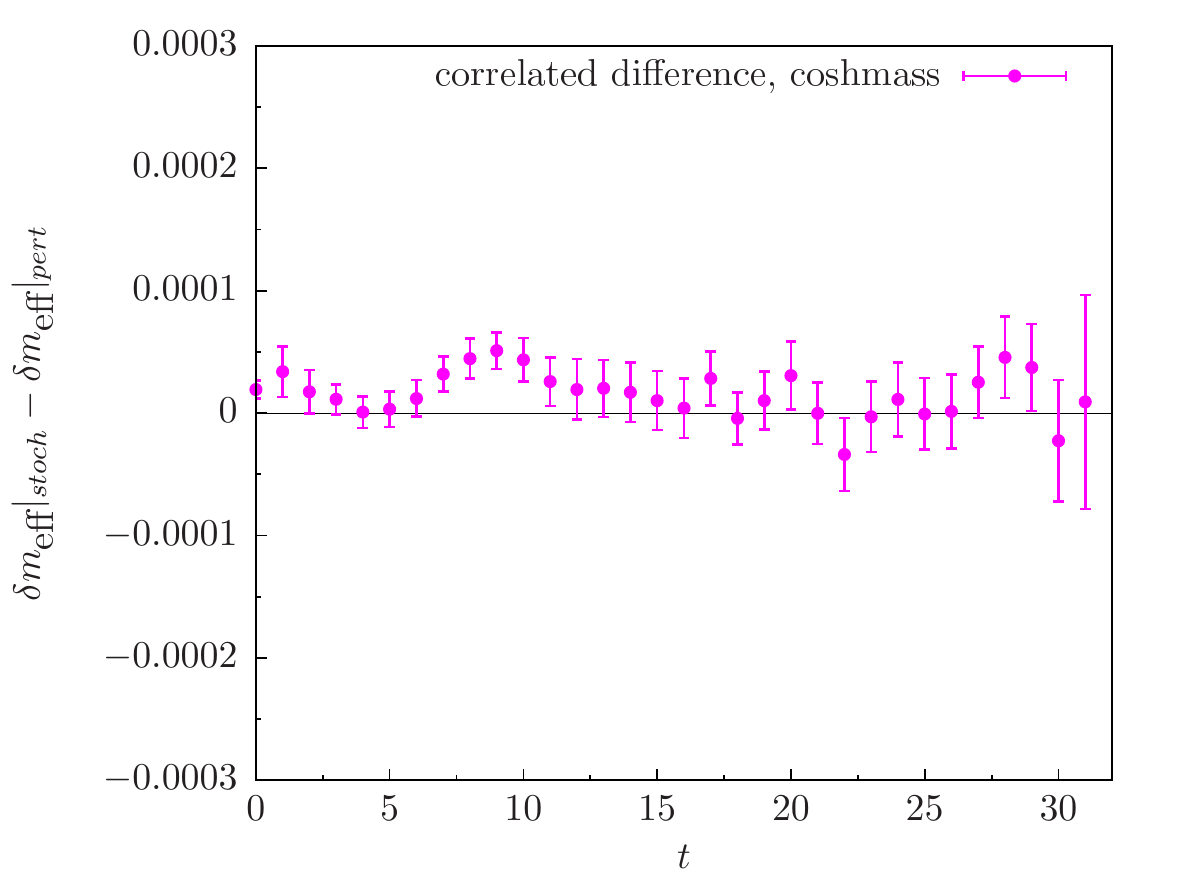}
\caption{The QED correction to the effective mass for a 
charged kaon is shown on the left. Blue circles show results from the 
stochsatic method, red squares results from the perturbative method. The plot 
on the right-hand side shows the correlated difference of both datasets.}
\label{fig:effmass_qed}
 \end{figure}

To compare the results from the stochastic and the perturbative methods, we 
determine the QED correction to the effective mass. For the stochastic data, we 
simply calculate the effective mass once for the two-point function including 
QED and once for the two-point function without QED and calculate their 
difference
\begin{equation}
 \delta m_\textrm{eff}(t) =  m^{\textrm{\scriptsize QED}}_\textrm{eff}(t) - 
m^0_\textrm{eff}(t)\,.
\label{eq:qed_effmass}
\end{equation}
For the perturbative method we make use of equation (\ref{eq:deltam}) and 
determine the QED correction to the effective mass as
\begin{equation}
\delta m_\textrm{eff}(t) = \frac{C_{\Oalpha}(t)}{ C_{0}(t)} - 
\frac{C_{\Oalpha}(t+1)}{ C_{0}(t+1)}\,.
\end{equation}
The left-hand side of figure \ref{fig:effmass_qed} shows the QED correction to 
the effective mass of a charged kaon obtained in both methods using the Feynman 
gauge for the photon fields. Blue circles show the results from the stochastic 
method, red squares the results from the perturbative method. The plot on the 
right-hand side of figure \ref{fig:effmass_qed} side shows the correlated 
difference 
of both datasets. Both datasets have been obtained on the same $87$ QCD gauge 
configurations using $16$ source positions for the stochastic and $4$ source 
positions for the perturbative method. For the $4$ source positions both 
datasets have in common we have used the same $\mathbb{Z}_2$ wall 
sources for the quark propagators. 
We find good agreement between the data from both methods. Note, that one 
expects both approaches to differ at $\Oalphasquare$, however, our data 
suggests, that these effects are smaller than the statistical precision.
\par 
For a comparison of the statistical precision of both methods, one has to take 
into account that the results have not been obtained at equal cost. For each 
quark flavor and source position the stochastic method requires three 
inversions ($q=0$, $+e$, $-e$). In the Feynman gauge (where only diagonal terms 
$\Delta_{\mu\mu}$ of the photon propagator contribute), the perturbative method 
requires $17$ inversions per quark flavor and source position to construct all 
the diagrams shown in figure \ref{fig:qeddiagrams}. Thus, for the result shown 
here, the computational cost is $3\times 16$ ($\#$ inversions $\times$ $\#$ 
source positions) for the stochastic and $17\times 4$ for the perturbative 
method. 
\par
\begin{figure}[h]
 \centering
 \includegraphics[width=0.48\textwidth]{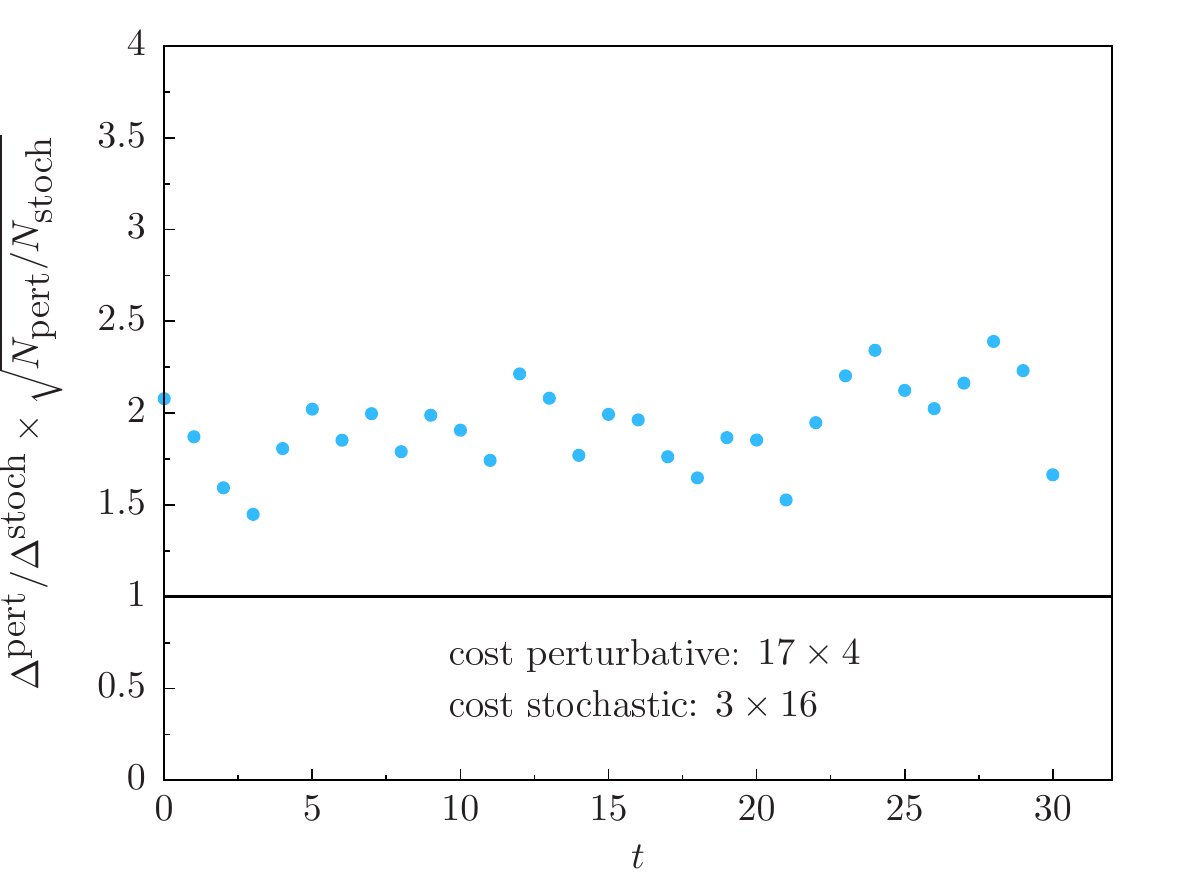}
  \includegraphics[width=0.48\textwidth]{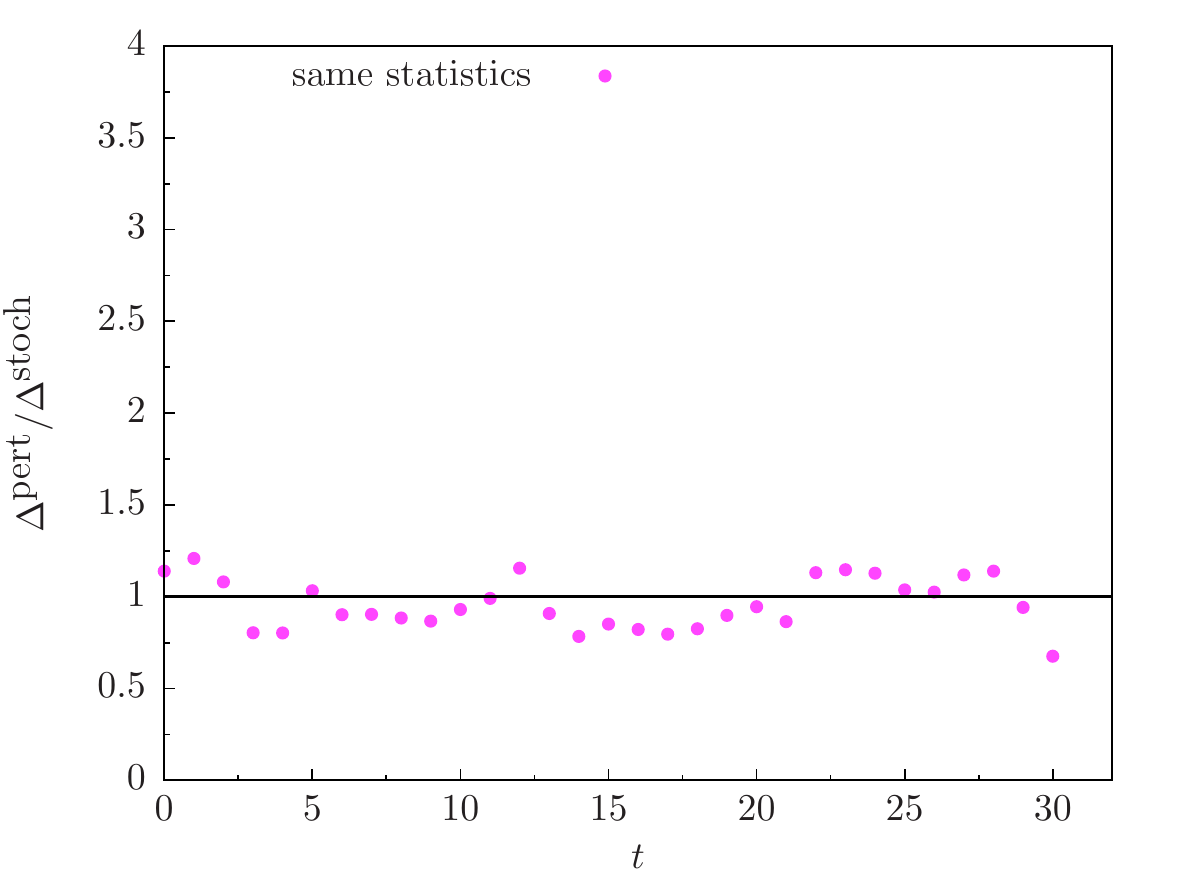}
\caption{Comparison of the statistical error on the QED correction to the 
effective kaon mass. On the plot on the left the errors for both methods are 
scaled with their respective computational cost. The plot on the right shows a 
comparison of statistical errors using the same amount of statistics ($4$ 
source positions).}
\label{fig:effmasscomp}
 \end{figure}

In figure \ref{fig:effmasscomp} we compare the statistical errors of the QED 
correction to the effective kaon mass for both methods. The left-hand side 
shows the ratio of the errors from the perturbative method over the stochastic 
method, both scaled with the numerical cost. The horizontal black line shows 
``$1$'', i.e. where both methods would be equal in statistical precision. 
However, we find that the error on the perturbative approach is about $1.5-2$ 
times larger than the error on the stochastic approach for the same 
computational 
cost. This might of course be a mass-dependent finding and differ for the 
physical values of the quark masses.
\par 
The right-hand side of figure \ref{fig:effmasscomp} shows a comparison of 
errors on the same set of statistics (using the same $4$ source positions in 
both data sets). Note, that this is not an equal cost comparison. Here we find 
both methods to have the same level of precision, indicating, that most of the 
noise originates from the QCD average.

\section{QED correction to the hadronic vacuum polarization}
\label{sec:hvp}
\subsection{Introduction}
The hadronic vacuum polarization (HVP) is the leading order hadronic 
contribution 
to the anomalous magnetic moment of the muon $a_\mu$. Since many years a 
deviation of about $3\sigma$ between the experimental value for $a_\mu$ and the 
Standard Model prediction persists, leaving room for new physics. Currently the 
most precise prediction for the HVP is obtained from the cross section of 
$e^+e^-\rightarrow$ hadrons \cite{hvp_ee_hadrons} with a precision of 
$\lesssim1\%$. In the past few years substantial progress has been made in 
calculating the HVP in Lattice QCD (see e.g.\ \cite{UKQCD_g2, Mainz_g2, 
ETMC_g2, HPQCD_g2,gm2strange}  )
aiming to be competitive with the prediction 
from $e^+e^-\rightarrow$ hadrons. 
\par
In this work we report a first attempt to estimate the QED corrections to the 
hadronic vacuum polarization. We calculate the HVP using a conserved vector 
current at the sink and a local 
current at the source
 \begin{equation}
 C_{\mu\nu}(x) = Z_V\,q_f^2 \left< V^c_{\mu}(x) V^\ell_{\nu}(0)\right>\,.
\end{equation}
Since one of the currents is non-conserved, we have to renormalize with the 
appropriate factor $Z_V$, which we take from a determination of the pion 
vector three-point function \cite{RBCUKQCD_Kl3}. The HVP tensor is constructed 
as
\cite{gm2strange}
\begin{equation}
 \Pi_{\mu\nu}(Q) = \sum_x e^{-iQ\cdot x} C_{\mu\nu}(x) - \sum_x C_{\mu\nu}(x)
 \label{eq:hvptensor}
\end{equation}
with a subtraction of the zero-mode $\sum_x C_{\mu\nu}(x)$ 
\cite{gm2strange,berneckermeyer}. For the HVP form factor we only consider 
contributions from (\ref{eq:hvptensor}) with vanishing spatial momentum $Q_i=0$ 
and estimate $\Pi(\hat{Q}^2)$ as
\begin{equation}
\Pi(\hat{Q}^2)  = \frac{1}{3} \sum_j \frac{\Pi_{jj}(Q)}{\hat{Q}^2}
\end{equation}
with $j=1,2,3$. From the HVP form factor the hadronic contribution to $a_\mu$ 
can be calculated as \cite{g2_lattice}
\begin{equation}
a_\mu^{\textnormal{HVP}} = \left(\frac{\alpha}{\pi}\right)^2  
\int\limits_0^\infty \textnormal{d} Q^2\, 
K(Q^2)\,\left[\Pi(Q^2)-\Pi(0)\right]\,.
\label{eq:amu}
 \end{equation}

\subsection{Results}

\begin{figure}[h]
\centering
\includegraphics[width=0.6\textwidth]{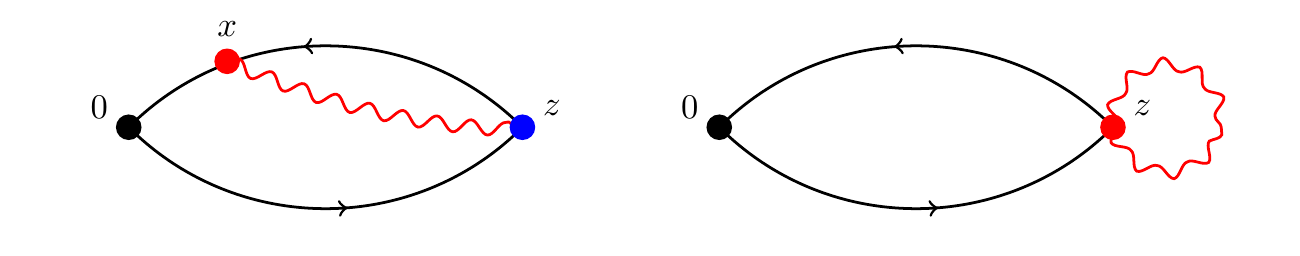}
 \caption{The additional diagrams for the QED correction to the HVP that occur 
when a conserved current is used at the sink. Red and blue vertices 
represent insertions of the 
conserved vector current and the tadpole operator, respectively.}
 \label{fig:qeddiagrams_hvp} 
\end{figure}

For the QED correction to the HVP calculated from the perturbative method, we 
have to consider two additional types of diagrams (cf figure 
\ref{fig:qeddiagrams_hvp}) further to the diagrams in figure 
\ref{fig:qeddiagrams}. These diagrams originate from a perturbative expansion 
in $\alpha$ of the conserved vector current at the sink. These two 
diagrams do not require any additional inversions.

\begin{figure}[h]
\centering
\includegraphics[width=0.48\textwidth]{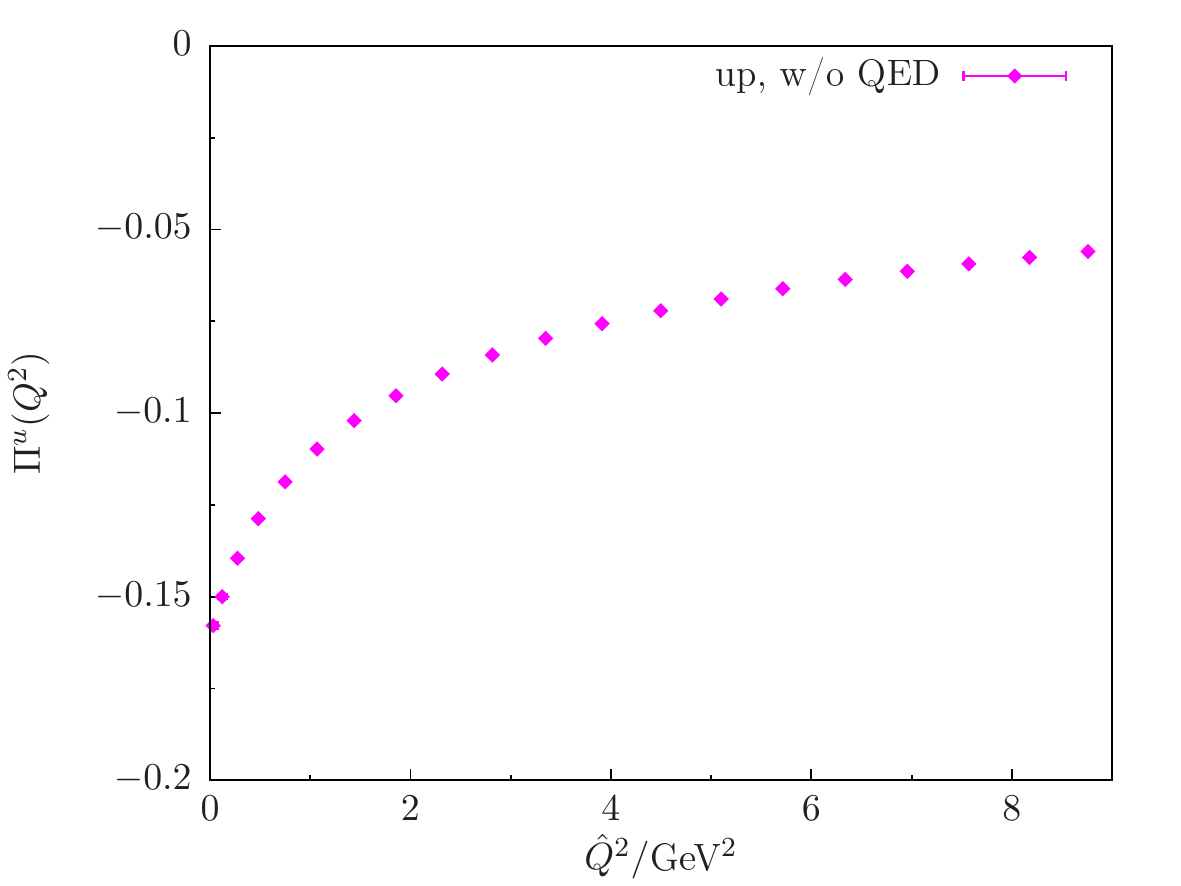}
\includegraphics[width=0.48\textwidth]{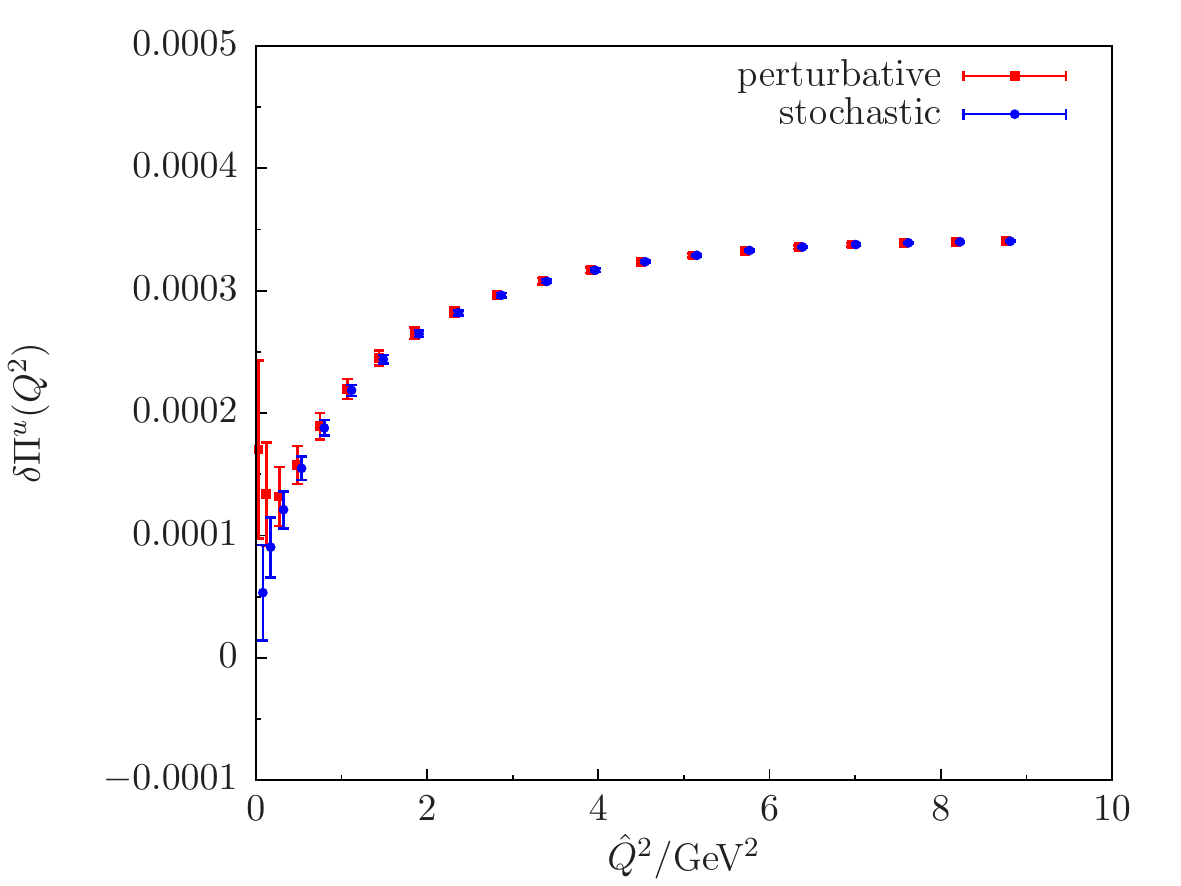}\\
\caption{The up quark hadronic vacuum polarization form factor 
$\Pi(Q^2)$ without QED is shown on the left. The plot on the right shows the 
QED correction to the HVP form factor for the u quark. 
Blue circles show data obtained using the stochastic method, red squares show 
data from the perturbative method.}
 \label{fig:hvp_qedcorr} 
\end{figure}

\begin{figure}[h]
\centering
\includegraphics[width=0.48\textwidth]{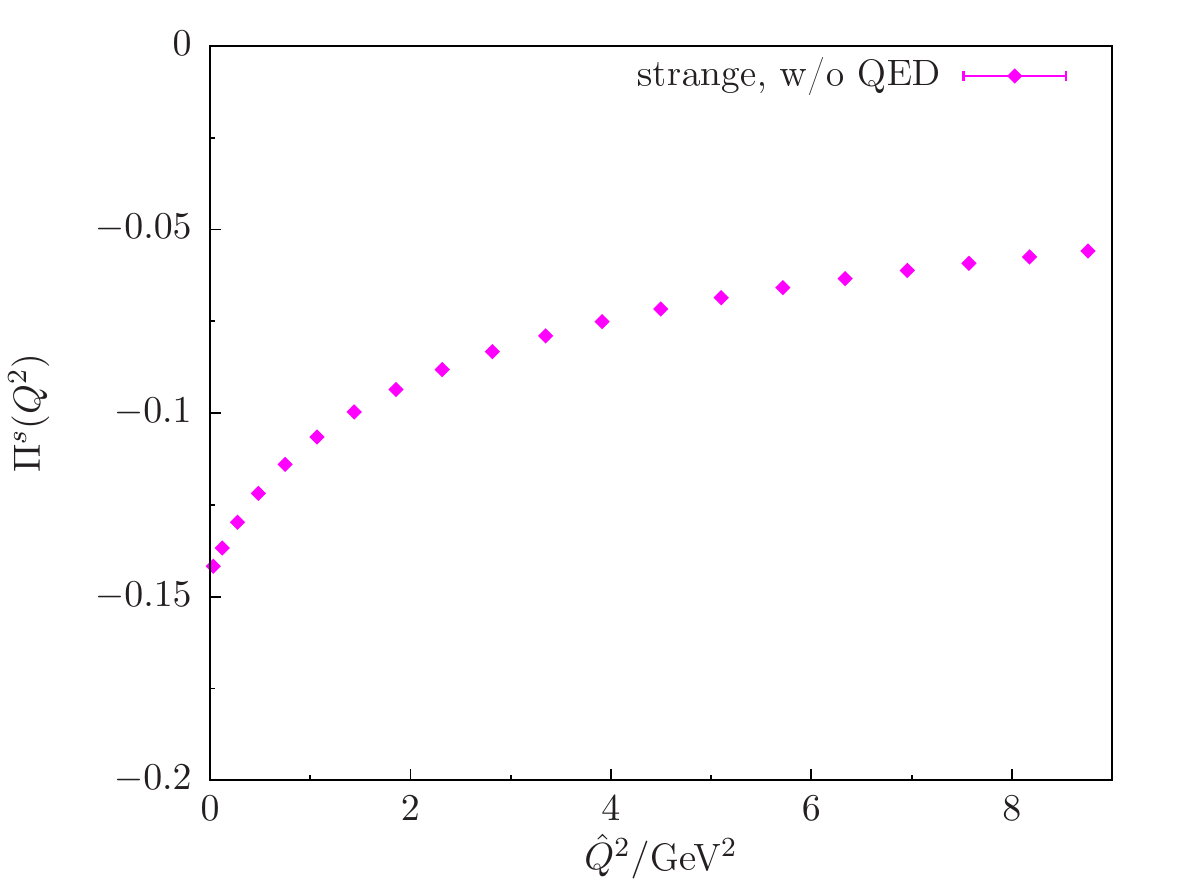}
\includegraphics[width=0.48\textwidth]{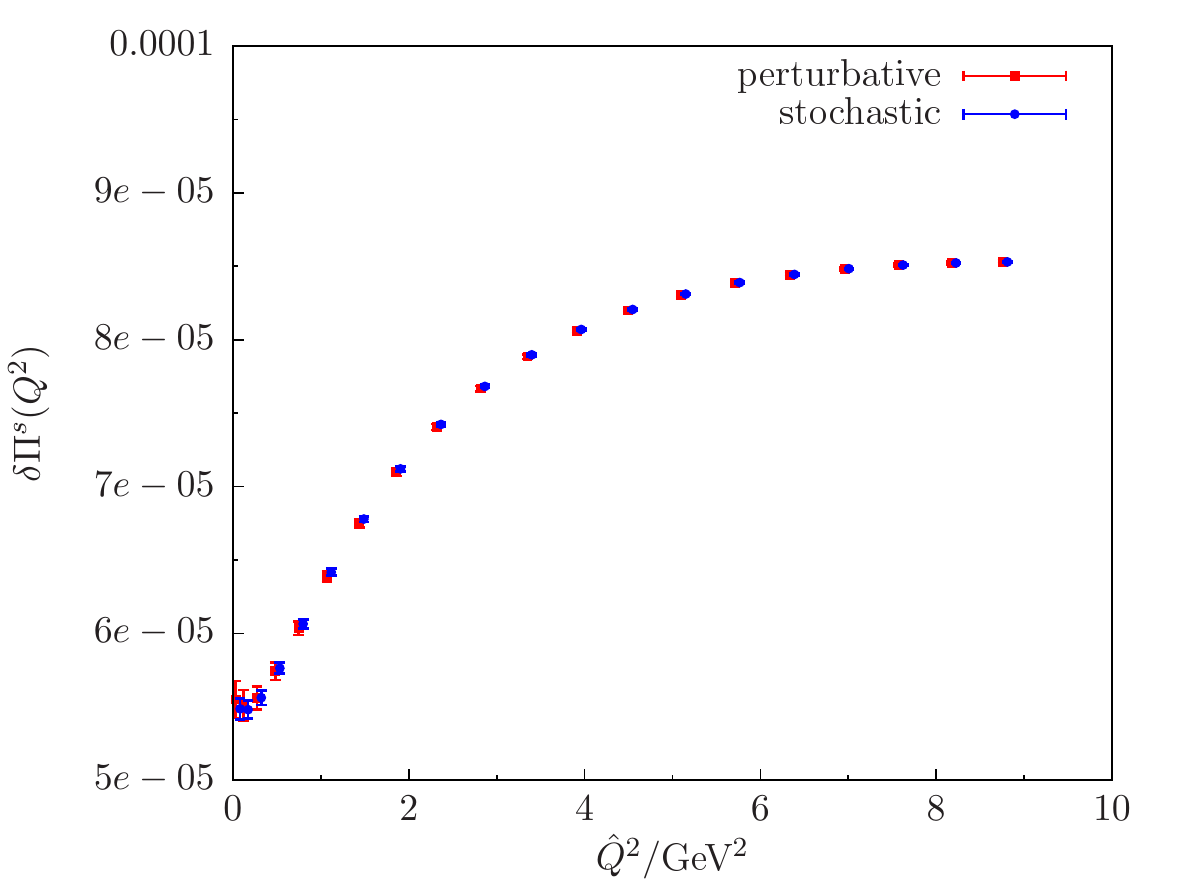}\\
\caption{The same as figure 8 for the strange quark.}
 \label{fig:hvps_qedcorr} 
\end{figure}

Figures \ref{fig:hvp_qedcorr} and \ref{fig:hvps_qedcorr} show results for the 
HVP form factor without QED (plots on the left) and the QED correction (plots 
on the right) for up and strange quarks, respectively. On the right-hand side 
plots, we show data obtaind from the stochastic method (blue circles) and the 
perturbative method (red squares). We find agreement between the results for 
the QED correction to the HVP from both methods. To obtain this consistency it 
is crucial to include the contact terms in figure \ref{fig:qeddiagrams_hvp} in 
the perturbative approach.

To get a first glimpse of the size of the QED correction to the anomalous 
magnetic moment of the muon we use \cite{berneckermeyer}
 \begin{equation}
 \hat{\Pi}(\hat{Q}^2) = \Pi(\hat{Q}^2) -\Pi(0) = 2\sum_z\, 
C_{jj}(z)\,\left[\frac{z_0^2}{2}-\frac{1-\cos(Qz_0)}{Q^2}\right]
\label{eq:hvprenorm}
\end{equation}
to renormalize the HVP form factor. We fill the momenta $Q^2$ with non-lattice 
momenta using a sine cardinal interpolation \cite{gm2strange} and integrate 
naivly using a trapezoidal rule to obtain $a_\mu$ according to equation 
\eqref{eq:amu}. However, in this way we are, at the current level of 
statistics, 
not able to resolve the QED correction to the renormalized HVP 
$\hat{\Pi}(\hat{Q}^2)$ in the low-$Q^2$ region which is most relevant for 
$a_\mu$. Thus, we obtain the QED corrections to $a_\mu$, which are consistent 
with 
zero within the statistical errors. Nevertheless, this still allows us to quote 
an upper limit for the QED corrections, which we find to be $\lesssim1\%$ for 
the up quark, and even smaller for down and strange, where the QED correction 
is suppressed by a factor $4$ compared to the up, due to the charge factors 
from the internal coupling of photons.
\par
In addition, one has to consider the QED correction to the multiplicative 
renormalization $Z_V$ for the local vector current in the HVP
\begin{equation}
 Z_V = Z_V^0 + \alpha Z_V^1\,,
\end{equation}
which yields at $\Oalpha$ a further QED correction to the HVP of the form $ 
Z_V^1\,\Pi^0(\hat Q^2)$ with $\Pi^0(\hat Q^2)$ the HVP without QED. We aim to 
include this correction to $Z_V$ in our future work.

\subsection{Comparison of statistical error}
\begin{figure}[h]
\centering
\includegraphics[width=0.48\textwidth]{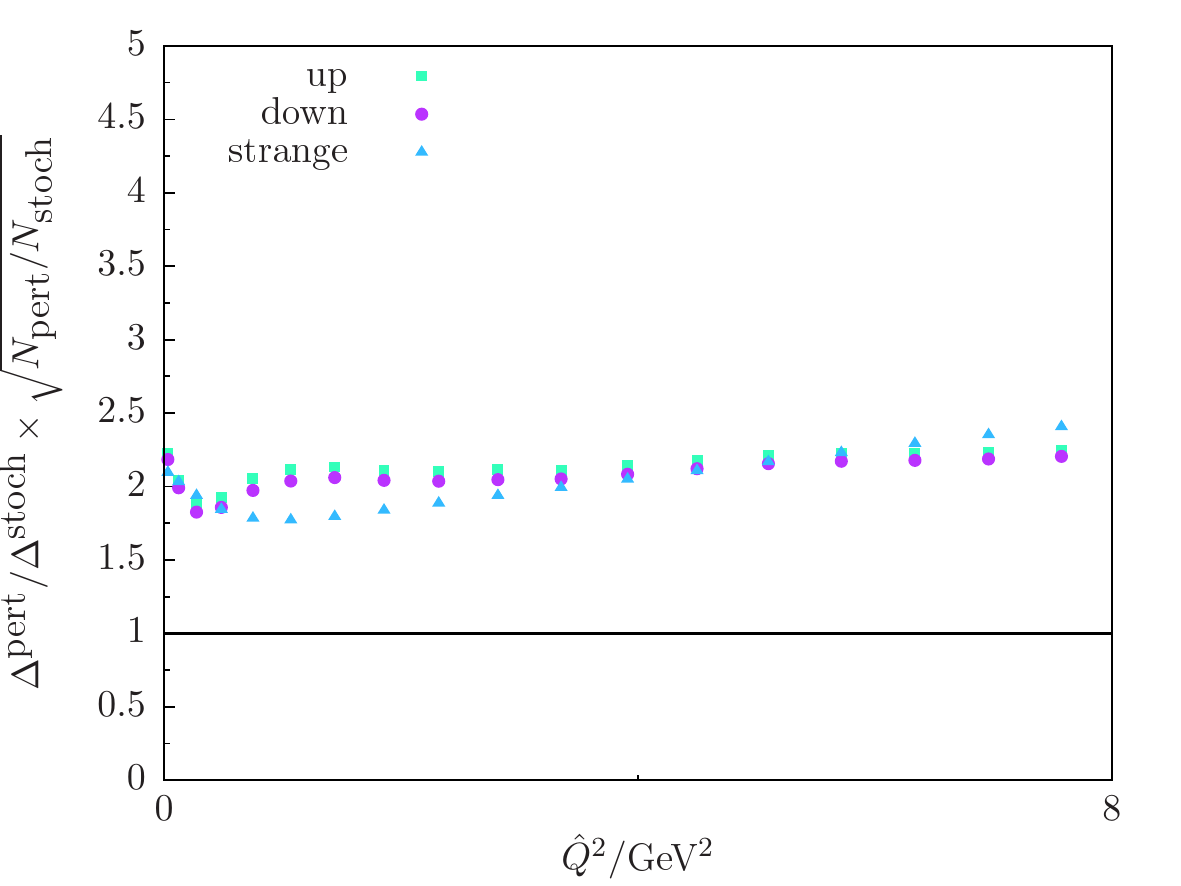}
\includegraphics[width=0.48\textwidth]{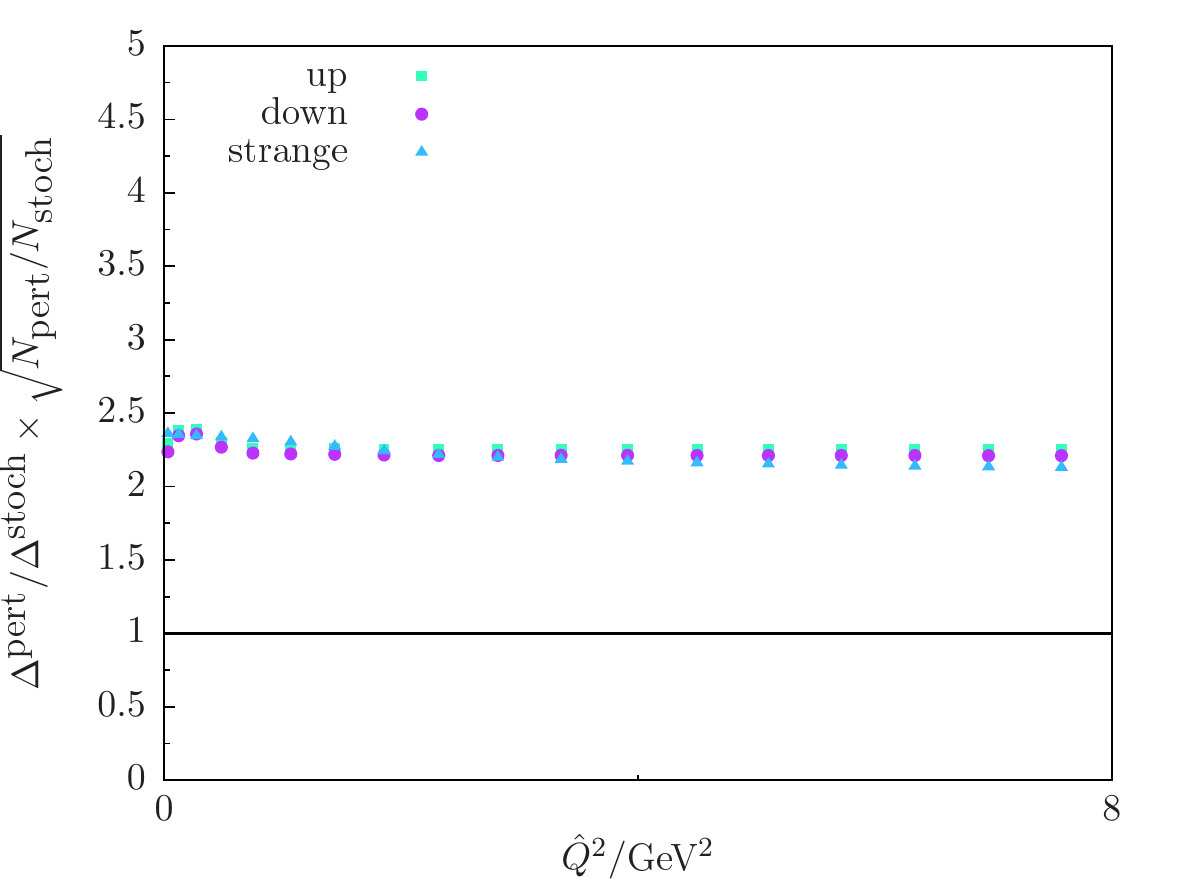}\\
\caption{Comparison of the statistical errors on the QED correction to the HVP 
from both methods for the HVP form factor (left) and the renormalized HVP 
(right). The plots show the ratio of the errors from the perturbative method 
over the stochastic method, both scaled with the numerical cost.}
 \label{fig:hvp_error} 
\end{figure}

In figure \ref{fig:hvp_error} we show the ratio of the statistical errors of 
the perturbative method over the stochastic method, scaled with the 
computational cost. The horizontal black line shows ``1'', i.e. where both 
methods would be equal in statistical precision. We find the stochastic method 
to give an $\approx2$ times smaller statistical error for the HVP form factor 
and $2-2.5$ times for the renormalized HVP for the same numerical cost (i.e.\ 
the same
number of inversions). We find this to be independent of the quark masses 
within the range studied here.

\section{Summary and Outlook}
\label{sec:conclusions}
We have calculated the QED corrections to meson masses and the hadronic vacuum 
polarization using two different approaches - a stochastic approach using 
$U(1)$ gauge configurations for the photon fields, and a perturbative approach 
by expanding the path integral in the electromagnetic coupling $\alpha$. We 
find agreement between the results from both methods, and with a 
previous independent determination \cite{RBCUKQCD_qed} of the QED corrections 
to the meson masses on the same gauge ensemble using a stochastic approach.
\par
Comparing the statistical errors on the QED corrections from both approaches 
for the meson masses and the HVP, we find the stochastic approach to yield a 
factor of $\approx2$ smaller errors for the same numerical cost, i.e. the 
number of inversions.
\par
We find the QED correction to the anomalous magnetic moment to be consistent 
with zero within the statistical errors, but at most of the order of $1\%$. 
Using a more sophisticated way to determine the additive renormalization 
$\Pi(0)$, e.g.\ using Pad\'e approximants, we might be able to 
resolve $\hat{\Pi}(\hat{Q}^2)$ in the low-$Q^2$ region. In addition, we are 
currently increasing statistics for the QED correction to the HVP, and 
investigating the use of all-mode-averaging \cite{ama}. 
\par
\begin{figure}[h]
 \centering
 \includegraphics[width=0.3\textwidth]{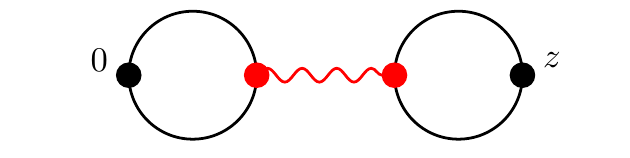}
 \caption{Disconnected diagram for the QED correction to the HVP.}
 \label{fig:disc_diagram}
\end{figure}

We plan to estimate 
the QED correction to the multiplicative renormalization for the local vector 
current $Z_V$, which results in a further QED correction of the form 
$Z_V^1\,\Pi^0(\hat Q^2)$. A more complete picture of the QED correction to the 
HVP will also require the inclusion of certain quark-disconnected diagrams, 
like the one shown in figure \ref{fig:disc_diagram}, which is not $SU(3)$ 
flavor suppressed, in contrast to its counterpart without QED, i.e. without a 
photon coupling the 
two quark loops.

\par
Another important issue that we will address in the future is a study of finite 
volume 
effects for the QED correction to the HVP. Finite volume corrections with 
photons in a finite box are expected to be significant, and it is therefore 
important to include them to obtain a precise result.

\section*{Acknowledgments}
This work has received financial support from the EPSRC Centre for
Doctoral Training in Next Generation Computational Modelling grant
EP/L015382/1. 
This work has received financal support from the  STFC 
Grant ST/L000296/1. 
The research leading to these results has received funding 
from the European Research Council under the European Union's Seventh Framework
Programme (FP7/2007-2013) / ERC Grant agreement 279757.
P.A.B. and A.P. are supported in part by UK STFC grant ST/L000458/1.
This work used the DiRAC Blue Gene Q Shared Petaflop system at the
University of Edinburgh, operated by the Edinburgh Parallel Computing
Centre on behalf of the STFC DiRAC HPC Facility (www.dirac.ac.uk).
This equipment was funded by BIS National E-infrastructure capital
grant ST/K000411/1, STFC capital grant ST/H008845/1, and STFC DiRAC
Operations grants ST/K005804/1 and ST/K005790/1. DiRAC is part of
the National E-Infrastructure.

\end{document}